\documentstyle[epsf,psfig,graphicx]{mn}

\def\x2{$\chi^{2}$}

\def\lunits{$\rm erg~s^{-1}$}
\def\funits{$\rm erg~cm^{-2}~s^{-1}$}
\def\cunits{$\rm cm^{-2}$}

\newbox\grsign \setbox\grsign=\hbox{$>$} \newdimen\grdimen \grdimen=\ht\grsign
\newbox\simlessbox \newbox\simgreatbox \newbox\simpropbox
\setbox\simgreatbox=\hbox{\raise.5ex\hbox{$>$}\llap
     {\lower.5ex\hbox{$\sim$}}}\ht1=\grdimen\dp1=0pt
\setbox\simlessbox=\hbox{\raise.5ex\hbox{$<$}\llap
     {\lower.5ex\hbox{$\sim$}}}\ht2=\grdimen\dp2=0pt
\setbox\simpropbox=\hbox{\raise.5ex\hbox{$\propto$}\llap
     {\lower.5ex\hbox{$\sim$}}}\ht2=\grdimen\dp2=0pt

\def\asca{{\it ASCA~}}

\def\xmm{{\it XMM-Newton~}}
\def\chandra{{\it Chandra~}}
\def\axj{AXJ0341.4-4453~}
\def\avx{$\rm A_V^X~$}

\begin{document}

\title[\xmm observations of an absorbed QSO]
{\xmm observations of an absorbed z=0.67 QSO:   
 no dusty torus  ?}

\author[Georgantopoulos et al.]
{ {\LARGE I. Georgantopoulos$^1$, A. Georgakakis$^1$, G.C. Stewart$^2$, 
 A. Akylas $^{1,3}$,} \\
 {\LARGE B.J. Boyle$^4$, T. Shanks$^5$, R.E. Griffiths$^6$} \\
$^1$ Institute of Astronomy \& Astrophysics, National Observatory of Athens, 
 I. Metaxa \& B. Pavlou, Penteli, 15236, Athens, Greece \\
$^2$ Department of Physics and Astronomy, University of Leicester, 
 Leicester LE1 7RH \\
$^3$ Physics Department University of Athens, Panepistimiopolis, 
 Zografos, 15783, Athens, Greece \\  
$^4$ Anglo-Australian Observatory, PO Box 296, Epping, NSW 2121, Australia \\
$^5$ Physics Department, University of Durham,  Science Labs, 
 South Road, Durham, DH1 3LE \\
$^6$ Department of Physics, Carnegie-Mellon University, 5000 Forbes Ave., 
 Pittsburgh, PA 15213, USA \\
}

\maketitle
\begin{abstract}
 We present \xmm observations of AXJ0341.4-4453,
 a mildly reddened $\rm A_V< 7$ QSO at a redshift of z=0.672. 
 The \xmm spectrum  shows a large obscuring column 
 $\rm N_H \sim 10^{23}$ \cunits corresponding to 
 $\rm A_V\sim 70$, in agreement with previous 
 results based on the lower limit of the  
 \asca hardness ratio. The X-ray spectrum 
 is represented by a 'scattering' model
 with $\Gamma\approx 2.0$ with the scattered power-law
 normalization being a few per cent of the hard component. 
 No FeK line is detected with a 90 per cent upper 
 limit on its equivalent width of $\approx$360\,eV. 
 The large discrepancy 
 between the column density observed in X-rays and that 
 inferred from the Balmer decrement can be explained 
 by dust sublimation near the nucleus. 
 Then, the X-ray and the 
 optical obscuration come from two different regions:
 the X-ray close to the accretion disk  while the optical 
 at much larger $>0.25$ pc scales. 
 
\end{abstract}


\newpage

\section{Introduction}
\asca and {\it BeppoSAX}  surveys have discovered several examples 
 of absorbed QSOs 
 beyond the local Universe (e.g. Boyle et al. 1998,  
 Georgantopoulos et al. 1999, Fiore et al. 1999, 
 Akiyama et al. 2000, Akiyama, Ueda \& Ohta 2002). 
   The X-ray spectra of 
 these QSOs present large amounts of photoelectric 
 absorption (typically $\rm N_H>10^{23}$ \cunits).
 Interestingly though, their optical extinction
 is disproportionately small.
 This is in stark contrast with the optical properties 
 of most  nearby Seyfert galaxies 
 which present large amounts of  
 absorption in X-rays (e.g. Goodrich et al. 1994),
 posing questions on the 
 physical conditions of the absorbing screen
 at higher redshift.  

 Here, we present \xmm observations of a moderate redshift 
 (z=0.672) QSO (\axj) detected in our deep \asca survey (Georgantopoulos 
 et al. 1997). The optical spectrum as well as the 
 \asca hardness ratio were presented by Boyle et al. (1998). 
 The optical spectrum, obtained in the wavelength range 
 4000-7000 \AA \  with the LDSS spectrograph in the Anglo-Australian 
 Telescope, shows only narrow lines. As some 
 of them are high excitation lines (e.g. [NeV] it was 
 immediately established that this is an AGN). 
 Subsequent spectroscopic observations obtained at the 
 CTIO by Halpern, Turner \& George (1999) 
 at longer wavelengths revealed a broad $\rm H_\beta$ line 
 clearly indicating that this is a broad-line (type-1) QSO.
 The excellent quality optical spectrum obtained by Halpern et al. 
 (1999) allowed the determination of the reddening from 
 the ratio of the Balmer lines: $\rm H_\gamma/H_\beta=0.17$ 
 yielding $\rm A_V=7$. They note however, that 
 the estimate on $\rm A_V$ is just an upper limit, 
 as the $\rm H_\gamma/H_\beta$ ratio could be affected 
 by collisional excitation, radiative transfer effects
 and contamination of $\rm H_\beta$ by $\rm Fe_{II}$.  
 Indeed, the $\rm MgII/H_\beta$ ratio  
 gives instead $\rm A_V=3.5$.  
  In contrast, the \asca hardness ratio gives a {\it lower limit} on the 
 column density of $\rm N_H>4\times 10^{22}$ \cunits which 
 corresponds to $\rm A_V \sim 30$ (Bohlin et al. 1978). 
 The high effective area of \xmm (more than an order of 
 magnitude higher than \asca at 1 keV) offers the opportunity 
 to study in detail the X-ray spectrum of this 
 object, expanding on the \asca results and 
 helping to explore the relation between 
 the obscuration at different wavelengths  
 beyond the local Universe.  

 Throughout this paper we adopt $\rm H_o=65
 km~s^{-1}~Mpc^{-1}$ and $q_o=0.5$.

\section{Data Analysis}
 The \xmm data  of QSO AXJ0341.4-4453 
 ($\rm \alpha=3h41m24.4s$;
  $\rm \delta=-44^{\circ}53^{\prime}2.0^{\prime\prime}$, J2000) 
 were obtained in 2002 August 26 as part of the
 Guaranteed Time program.  The EPIC (European Photon Imaging Camera;
 Str\"uder et al. 2001 and Turner et al. 2001) cameras were operated
 in full frame mode with the thin filter applied.    

 The Data have been cleaned and  processed using the  Science Analysis 
 Software ({\sc sas} 5.3). The event files for the PN and the two MOS
 detectors were produced using the standard reduction pipeline
 (Watson et al. 2001). The event files were screened for high
 particle background periods resulting in a good time interval of 
 26122\,s. Events corresponding to patterns 0--4 and 0--12
 have been included in the analysis of the PN and MOS data
 respectively.    
  
 \axj is observed on-axis. Its X-ray centroid lies at about
 1\,arcsec away from the optical counterpart. For the X-ray spectral
 analysis an extraction radius of 12\,arcsec is adopted.  We create an
 auxiliary file using the {\sc SAS} task {\sc arfgen} to take into account
 both instrumental effects (e.g. quantum efficiency, telescope
 effective area, filter transmission)  and the fraction of light
 falling outside the 12\,arcsec extraction radius. The spectral
 response file is created using the task {\sc rmfgen}. The source
 spectrum is grouped, using the {\sc ftool} {\sc grppha} task to give
 a minimum of 15 counts  per bin to ensure that Gaussian
 statistics apply.  The source spectrum is fitted using {\sc xspec}
 v11.0. The Galactic column density is $1.7\times 10^{20}$\,\cunits \,  
 (Dickey \& Lockman 1990).  We fit the data in the energy range
 0.3-10\,keV where  the sensitivity of \xmm is the highest. The errors
 quoted  below correspond to the  90 per cent confidence level.     
 
 As a first approximation, a power--law model is fit to the data,
 yielding the intrinsic absorption column density, $\rm N_H$, and 
 the power--law photon-index $\Gamma$.  The spectral fit results 
 are summarized
 in Table 1 and are shown in Figure 1. An acceptable fit
 ($\chi^2_\nu=24.0/23$) is obtained for a rest frame column density
 $\rm N_H\sim 10^{23}$\,\cunits, consistent with the lower limit
 obtained by previous \asca  observations ($\rm N_H>4\times
 10^{22}$\,\cunits).     
 Assuming the standard Galactic $\rm A_V/N_H$ factor the above column
 density corresponds to an optical absorption $\rm A_V\sim 70$ 
 (Bohlin et al. 1978) much
 higher than the estimated amount of dust extinction   $\rm A_V<7$
 based on Balmer decrement measurements (Halpern et al. 1999). 

 It is clear from Figure 1, that an excess at low energies 
 over the {\it absorbed} power--law
 model is present. Therefore, we investigate whether a more complex
 model could  compensate for this spectral feature. In particular, the
 'scattering' model is examined. This assumes a thick  screen of
 absorbing material which swamps the soft X-rays, while the more
 energetic hard X-rays can penetrate through. In this picture the
 soft-band emission arise from scattering on a pure electron
 medium. In practice this model is realized by a two component
 power--law: the first power--law is absorbed by the intrinsic column  
 density, while the second is absorbed only by the  Galactic
 column. An additional constrain is that  the scattered soft power--law
 should  have the same slope as the hard component.  This  standard
 model appears to apply  to most obscured Seyfert galaxies 
 (e.g. Turner et al. 1997). The results of the spectral fits are also
 presented in Table 1. We obtain a reduction in $\chi^2$ of $\Delta
 \chi^2\approx6.1$ which  is statistically significant at the
 $\approx99$  per cent confidence level for one additional parameter
 according to the F-test (Bevington \& Robinson 1992). 

 We also check whether the X-ray spectrum presents an FeK line at a
 rest-frame of 6.4\,keV. This line is often detected in obscured
 Seyfert galaxies having equivalent widths $>200$\,eV (e.g. Turner et
 al. 1997). This is believed to be produced through transmission
 of the X-ray radiation in a dense cold medium. When we add a {\it
 narrow}  Gaussian line (with $\sigma=0.01$\,keV much smaller than
 EPIC's spectral resolution) we obtain $\Delta \chi^2\approx 1.4$
 which does not represent a statistically  significant change. We find
 a 90 per cent upper limit on the equivalent width of  $\sim$360\,eV
 (rest-frame). This is roughly consistent with the equivalent width of
 $\sim 300$\,eV  predicted for an  obscuring screen of
 $10^{23}$\,\cunits \, assuming spherical geometry (Leahy \& Creighton
 1993).    
 
 Moreover, we check for long-term variability between the \asca and
 \xmm epoch.  Using the single power--law model, the source flux  
 and luminosity in the 2-10\,keV band is $8\times 10^{-14}$\,\funits \, 
 and $8\times 10^{43}$\,\lunits \,  respectively. Therefore \axj presents
 no variability  when compared to the  \asca observation:  $\rm f_{2-10
 {\rm keV}} \approx  8.6 \pm 2.3 \times 10^{-14}$\,\funits, using   
 again the best-fit \xmm spectral parameters in the single
 power--law case $\Gamma=2.05$,  $\rm N_H=10^{23}$\,\cunits.
 Finally, we check for short term variations in the \xmm light curve
 by binning the data in 1\,ks bins. A constant line is an acceptable
 fit to the flux of the binned data ($\chi^2=37.2/29$) and hence, the 
 observations do not reveal short-term variability on timescales of
 hours.   
    
\begin{table} 
\caption{Best fit parameters}
\begin{tabular}{ccccc}

\hline 
Model& $\rm N_{H}^1 $ & $\Gamma$ & $f^2$ &\x2/dof  \\
\hline 

Single Power--Law &   $0.9^{+0.3}_{-0.4}$ & $2.06^{+0.56}_{-0.37}$ & - & $24.0/23$ \\

Scattering &  $1.0^{+0.4}_{-0.3}$ & $2.05^{+0.28}_{-0.29}$ & 
         $0.035^{+0.035}_{-0.030}$ & $17.9/22$ \\

\hline 

\end{tabular} 

$^{1}$ Neutral column density $\times 10^{23}$ \cunits \\
$^{2}$ Normalization of the soft relative to 
 the hard power--law \\

\end{table}

\begin{figure} 
\centerline{\psfig{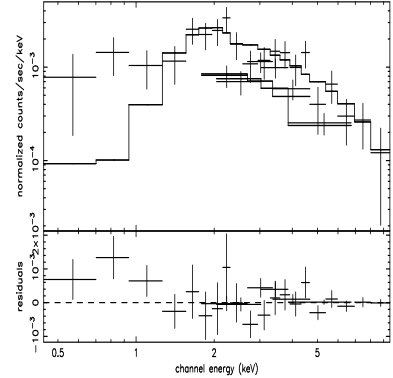}} 
\caption{The single power--law spectral fit to the PN 
 and MOS data 
 (upper panel) together with the residuals
 (bottom panel). A soft excess at low 
 energies ($<2$\,keV) is clearly present}
\end{figure}

\section{DISCUSSION} 

 The \xmm observations show that 
 \axj presents an X-ray spectrum typical of Seyfert-2 galaxies 
 in the local Universe with the     
 absorbing column being high ($\rm N_H\sim10^{23}$ \cunits).  
 However, there is a large  discrepancy between the 
 obscuration estimates derived
 from the optical spectrum (using the Balmer decrement) and the X-ray
 column density. Several QSOs in the sample of Akiyama et
 al. (2000) as well as a couple  of QSOs in the {\it BeppoSAX} HELLAS
 sample  of Fiore et al. (1999) exhibit the same behaviour.  
   
 Granato et al. (1997) argue that dust sublimation can explain the
 discrepancy in the optical and X-ray $\rm A_V$ estimates in these objects. 
 In more detail 
 dust is destroyed  in the presence of the strong AGN radiation field
 in the vicinity of the sublimation radius estimated as $\rm r = 0.5
 \times L_{46}^{1/2}$\,pc, where $\rm L_{46}$ is the intrinsic UV luminosity 
 in  units of  $10^{46}$\,\lunits (Granato et al. 1997). 
 In the case of \axj we estimate $\rm L_{46}$ by extrapolating the X-ray
 power--law to the UV.  We assume an X-ray--to--optical spectral index
 $\alpha_{ox}=1.6$ typical of QSOs (e.g. Schmidt \& Green 1986) to convert 
 from X-ray
 2\,keV flux density to 2500\,\AA \ flux density. To estimate the UV flux
 between 1000--3500\,\AA \ we adopt a power--law spectral energy
 distribution at the UV of the form $f_\nu\sim\nu^{-0.5}$. We find
 that dust cannot exist within a radius of $\approx0.25$\,pc. 

 Maiolino, Marconi \& Oliva (2001) argue against the dust
 sublimation scenario on the basis of the observed emission--line
 properties.  They propose an alternative model suggesting that the
 high densities in the nuclear regions of AGNs favor dust coagulation
 resulting in a dust distribution dominated by large dust grains.
 Detailed modeling shows that such a dust grain distribution can
 effectively explain the observed lower values of reddening compared
 to those expected from the hydrogen column density measured in
 X-rays assuming a standard Galactic dust--to--gas ratio (Maiolino,
 Marconi \& Oliva 2001). Observational and theoretical evidence also
 support the above picture of `anomalous' dust properties in the AGN
 nuclear regions (Maiolino et al. 2001; Clavel et al. 2000). 
 
 In a recent paper Weingartner \& Murray (2002) using the sample of
 Maiolino et al. (2001)  report the absence of any correlation
 between  the optical reddening and X-ray column density. They
 interpret this as evidence that the optical obscuration  and the
 X-ray photoelectric absorption  occur in two distinct
 media. Consequently, they argue that the basic assumption of Maiolino
 et  al. (2001) that the optical  and photoelectric absorption are
 taking place on the same medium (i.e. the torus) might not hold. 
 In their picture X-rays are absorbed by dust-free (or large dust
 grain) material just off the torus and/or the accretion disk, while
 the optical extinction occurs further out from the nucleus on a
 medium with nearly Galactic properties. They also claim that several
 narrow-line AGN in the sample of Maiolino et al. (2001) have
 erroneous estimates of the dust extinction due dust lanes in the host
 galaxy disks not associated with the torus.

 Here, we re-examine the relation between the ratio of the X-ray to
 optical obscuration  in the sample of Maiolino et al. (2001),
 excluding the eight   objects for which  Weingartner \& Murray (2002)
 raised criticism: NGC\,1365, MCG\,5-23-16, NGC\,5506, NGC\,2992,
 IRAS\,13197-1627,  IRAS\,0518, NGC\,526a, Mrk\,231. We also  exclude  
 NGC\,4639 for which the error of the  measurement of the column
 density is very large. We further add to the sample two high redshift
 QSOs: RXJ13334+001 ($z=2.35$; Georgantopoulos et al. 1999)  and 
 AXJ08494+4454 ($z=0.9$; Akiyama et al. 2002). Note that
 our estimate of the reddening in the case of RXJ13334+001 is only
 approximate. The $\rm H_\beta$ line is not detected and hence the
 Balmer decrement method can only estimate a lower limit 
 $\rm A_V >3$. However, as the $\rm H_\alpha$ line is detected, the reddening 
 cannot be much higher. We estimate that a reddening of $\rm A_V \sim 10$
 would render $\rm H_\alpha$ undetectable to the sensitivity of the
 spectroscopic observations of Georgantopoulos et al. (1999). Note
 also  that although the QSO presented in the present paper was 
 included in the Maiolino et al. (2001) sample,  only an upper  limit
 on the ratio of the optical reddening  to the X-ray column density
 was available at the time (a lower limit on $\rm N_H$ 
 was determined by \asca). Here, we use $\rm A_V=7$ for this object.    
 
 In Figure 2, we plot the ratio of the X-ray to optical column  
 $\rm R=A_V^O/A_V^X$ as a function of redshift where \avx$\rm
 =2.2\times10^{21}~ cm^{-2}/N_H$ (Gorenstein 1975).  It should be
 noted that redshift is interchangeable with luminosity in Figure 2, ie
 the most X-ray luminous objects are also at higher $z$. The
 two lowest redshift objects (NGC\,5033, M\,81) 
 apparently have an optical to X-ray obscuration 
  higher than  that predicted on the basis 
 of the Galactic dust--to--gas ratio ($\rm R=1$). 
 These two objects have the lowest luminosities: 
 $\rm L_x < 1.2\times10^{41}$\,\lunits (see Table 1 of Maiolino et
 al. 2001). The rest of the sample has $R$ values lower than unity
 while there is no correlation with redshift. Under the dust
 sublimation scenario of Granato et al. (1997) for a source with  
 $\rm L_X>10^{42}$\,\lunits we estimate a dust sublimation radius of
 $\rm r>0.04$\,pc.    
 
 Recent X-ray monitoring  observations show that the size of the
 X-ray absorbing medium in several nearby Seyfert-2 galaxies 
 (Risaliti et al. 2002) is roughly  of the order of the Broad
 Line Region, $<0.03$\,pc. Interestingly, for the low luminosity
 AGNs in the Maiolino et al. sample ($\rm L_X\sim10^{41}$ \lunits) 
 we estimate a
 sublimation radius of $\rm r\sim0.01$\,pc and therefore for these systems
 dust does exist within the region that produces the photoelectric
 X-ray absorption. Hence, for these two systems it is probably not
 surprising that they do show large optical reddening. 
 For more luminous QSOs, $\rm L_X> 10^{42}$\,\lunits, the 
 sublimation radius is  $\rm r>0.04$\,pc and hence, the region of the
 X-ray absorbing medium is probably free of dust. For these systems the optical
 obscuration is taking place further out from the
 nuclear regions, presumably at pc-scale 
 or even larger distances (e.g. parent galaxy dust lanes). 
 The evidence above is in
 better agreement  with the Weingartner \& Murray (2002) picture where
 the optical and X-ray absorption are taking place on two distinct
 media. Additional constraints on the location of the 
 X-ray absorbing medium come from its ionization state. 
 If the absorbing screen is too close to the nucleus, 
 having a low density,  it will be fully ionised.    
 From the definition of the ionization constant 
 ($\xi=L/n_e~r^2$), we find that at a distance of 0.02 pc 
 and for the unobscured  luminosity of \axj,
 $\rm L_{0.1-2 keV}=5\times 10^{44}$ 
 \lunits,  the ionization constant is relatively low 
 ($\xi=1$) only for densities
 as high as $\rm n_e > 10^{11} cm^{-3}$. 

 However, one should be cautious about this interpretation because of
 selection biases in the Maiolino et al. (2001) sample. In particular,
 these authors argue that their sample is biased against AGNs 
 with enough optical obscuration to completely extinguish the broad
 lines. Therefore, the presence of AGNs with both 
 large amounts of X-ray obscuration and large optical extinction 
  cannot be excluded. The discovery of at least two type-2 QSOs (Stern
 et al. 2002, Norman et al. 2002) in \chandra surveys 
 suggests that this class of 
 sources does exist. Nevertheless, it is possible under our interpretation
 and also that of Weingartner \& Murray (2002) that the extinction of
 the broad lines in these systems is taking place on a medium external
 to the nuclear regions. For example, if these nuclei reside 
 in edge-on galaxies, this could explain the large optical reddening.   

 In conclusion, the large discrepancy between 
 the X-ray absorption and the optical reddening in \axj 
 cannot be explained by the standard 'dusty torus' 
 model where {\it both} the X-ray and the optical obscuration 
 are associated with an obscuring screen outside the BLR 
 at roughly pc-scale distances.  
 We suggest instead, a model where the X-ray 
 and the optical obscuration take place in two distinct media
 similar to that of Weingartner \& Murray (2002).  
 In particular, the X-ray absorption must take place close to the nucleus. 
 This region is dust--free in the more luminous  (and 
 thus high redshift) AGN as dust sublimates. 
 Then the small amount of  optical reddening  observed 
 in \axj comes from a 
 region further away from the nucleus ($>0.25$ pc) 
 where dust remains intact.

\begin{figure} 
\centerline{\psfig{figure=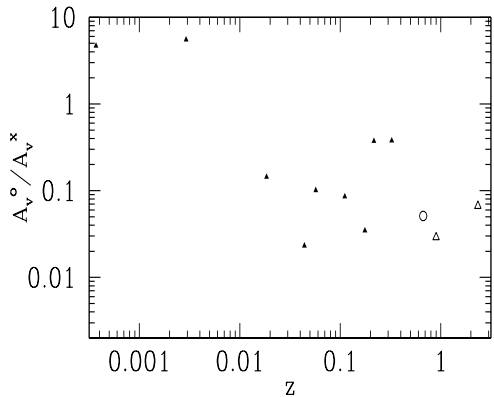,width=7cm,height=7cm,angle=270}} 
\caption{The ratio of the optical to the X-ray 
 obscuration ($\rm A_V^O/A_V^X$) as a function of redshift  
 for the sample of Maiolino et al. (2001) (filled triangles).  
 \axj (open circle) and two high redshift QSOs
  RXJ13334+001 and AXJ08494+4454 (open triangles); 
 see text for details}
\end{figure}

\section{acknowledgments}
 We thank the referee Prof. J. Halpern for useful comments and suggestions.  
 This work is jointly funded by the European Union and the Greek Goverment 
 in the framework of the programme 'Promotion of 
 Excellence in Technological Development and Research',  
 project 'X-ray Astrophysics with ESA's mission XMM'.

\end{document}